\documentclass[superscriptaddress,aps,epsfig]{revtex4}
\usepackage{graphicx}
\usepackage{epstopdf}
\usepackage{latexsym}
\usepackage{amssymb}
\usepackage{textcomp}
\usepackage{amsmath}
\usepackage[center]{subfigure}
\usepackage{feynmp}
\usepackage{slashed}
\usepackage{pdfpages}
\begin{document}
\renewcommand\thesection{\arabic{section}}
\renewcommand\thesubsection{\thesection.\arabic{subsection}}
\renewcommand\thesubsubsection{\thesubsection.\arabic{subsubsection}}
 \newcommand{\bq}{\begin{equation}}
 \newcommand{\eq}{\end{equation}}
 \newcommand{\bqn}{\begin{eqnarray}}
 \newcommand{\eqn}{\end{eqnarray}}
 \newcommand{\nb}{\nonumber}
 \newcommand{\lb}{\label}
\newcommand{\GeV}{\ensuremath{\,\text{sin}}\xspace}
\title{A Homogeneous and Isotropic Universe in Lorentz Gauge Theory of Gravity}

\author{Ahmad Borzou}
\email{ahmad_borzou@baylor.edu}
\affiliation{Physics Department, Baylor University, Waco, TX 76798-7316, USA}
\affiliation{Department of Physics, Isfahan University of Technology, Isfahan, 84156-83111, Iran}
\author{Behrouz Mirza}
\email{b.mirza@cc.iut.ac.ir}
\affiliation{Department of Physics, Isfahan University of Technology, Isfahan, 84156-83111, Iran}

\date{\today}

\begin{abstract}
Lorentz gauge theory of gravity was recently introduced. We study the homogeneous and isotropic universe of this theory.
It is shown that some time after the matter in the universe is diluted enough, at $z \sim 0.6$, the decelerating expansion shifts  spontaneously to an accelerating one without a dark energy. We discuss that Lorentz gauge theory puts no constraint on the total energy content of the universe at present time and therefore the magnitude of vacuum energy predicted by field theory is not contradictory anymore. It is demonstrated that in this theory the limit on the number of relativistic particles in the universe is much looser than in GR. An inflationary mechanism is discussed as well. We show that the theory, unlike GR, does not require the slow-roll or similar conditions to drive the inflation at the beginning of the universe. 
\end{abstract}

\maketitle

\section{Introduction}
Recently a Lorentz gauge theory of gravity (LGT) has been proposed in which metric is not dynamical \cite{Borzou2016,Borzou2016_2}. The theory is brand new and detailed studies are needed in every area. 
It is already shown that the theory has a predictable high energy behavior. Also, the Schwarzschild metric is proved to be an exact vacuum solution of LGT and consequently all of the solar system experimental tests of GR are already passed by LGT as well. The linear version of the theory is fully consistent with Newtonian gravity and Newton's gravitatinal constant is shown to be an effective coupling constant that naturally emerges if a very small length is integrated out of full LGT. This suggests that there should be no large hierarchy between the dimensionless coupling constant of LGT and that of the other three forces. Moreover, it is shown that in LGT there is no potential that linearly grows with the distance from the source, which is an unknown type of potential being observed in Poincare gauge theories (PGT). The de Sitter space-time is another exact vacuum solution of LGT, suggesting that there may be no need for dark energy to explain the late accelerating expansion of the universe. 

It is useful to mention that PGT also have an interesting explanation for the acceleration of the universe which is different from what we present in the current work. It is shown that in PGT there are two problem-free propagating modes related to torsion, scalar torsion modes. It is moreover shown that one of the two modes, in a homogeneous and isotropic universe, can provide an oscillating force that makes the universe accelerate for some time and decelerate for some other time and we happen to live in the acceleration mode today. For a review of the subject see the work of Nester and his collaborators and also the references provided by them \cite{Nester_}. The main difference between the mentioned work and the one being presented in this paper is that Nester's study was done within PGT while we are working within LGT and the two have totally different dynamics built on the same kinematics. In the former, this is the torsion that provides the force for the acceleration of the universe while in the latter torsion is assumed zero. 

Our goal in the present paper is to investigate whether LGT can provide a cosmological picture consistent with the current experimental data, that are summarized next. 
Analysis of the Sloan Digital Sky Survey indicates that our universe is homogeneous in scales larger than a few hundred light years \cite{Yadav2005}. Also, a study of the fluctuations in the temperature of the comic microwave background shows that the universe is isotropic. The existence of the CMB itself indicates that the universe is expanding. The expansion is proved to be accelerating at present time \cite{SupernovaCosmologyProject,SDSS2014,Planck2016A13}. 
The universe should be older than its contents and therefore its age is determined to be $\gtrsim$ 13 billion years through multiple observations \cite{WMAP2013_20,Hansen2002_1, Gratton2003}. 
Another important parameter in a cosmological model is the spatial geometry of the universe. Multiple experiments are pointing out that our universe is spatially flat, for example see \cite{PlanckCol,Larson2011}. 
The matter content of the universe can be divided into relativistic and non-relativistic categories. 
Except massless photons, the rest of matter become non-relativistic at some point. 
Neutrinos are perhaps non-relativistic today, depending on their masses that are not known yet, but since they were relativistic when they went out of equilibrium with the rest of matter, their temperature falls like relativistic matter. Consequently, their density should not differ much from that of photons. The observed value of the CMB temperature indicates that both photons and neutrinos have a density of about $10^{-4}$ of the critical density at present time. 
Observation of rotation curves of spiral galaxies indicate that either the Newtonian gravity or the standard model should be modified. Observation of the gravitational equipotential surfaces of the Bullet cluster strongly disfavors a modified Nowtonian gravity \cite{Clowe2006} and indicates that the non-relativistic matter today is made of baryonic and dark matter. Multiple independent observations  suggest that the baryon density today is only 0.04 of the critical density \cite{Rauch1997,Pryke2001}. Observations that are sensitive to the ratio of baryon to the whole matter suggest that the ratio is one to five and consequently the non-relativistic matter today is $\sim 0.3$ of the critical density \cite{Percival2001}. There is one other type of energy in the universe whose existence is confirmed experimentally, vacuum energy \cite{Casimir1948,Sparnaay1957}. Quantum field theory has a prediction for the magnitude of the vacuum energy that together with GR severely contradicts the observations \cite{Weinberg1989}. This is one of the major problems in physics that has not been solved yet. 
In this paper we do not study some very important observations in modern cosmology like the CMB anisotropies, light element abundances, large scale structures, etc. These are left for future works. 

In the following we first review the basics of LGT. Next, we derive the differential equations assuming a Friedmann Lemaitre Robertson Walker metric and find two exact solutions for them. Next, based on the observations discussed above, we set the initial values at present time and numerically solve the equations backward in time. The results are compered with the observations and their consistencies are checked. Finally a conclusion is drawn in the end.

\section{Lorentz gauge theory of gravity}
\lb{LGTsec}

Given the successes of the standard model as a gauge theory, one my wish to construct a similar theory for gravity. The construction of such a theory began with the work of Utiyama \cite{PhysRev.101.1597}. The concepts were further elaborated by the studies of Sciama and Kibble \cite{RevModPhys.36.463, Kibble1961}. Very good reviews of the subject can be found in 
\cite{Hehl19951, Blagojevic:2013xpa, IVANENKO19831, RevModPhys.48.393, Obukhov:2006gea, Shapiro2002113, VonDerHeyde1975, ASNA:ASNA2103020310, Obukhov1989, BAEKLER1987800, Vereshchagin:1999jv, Kuhfuss1986}. 
Assuming no parity violation, the most general form of the Lagrangian for gauge theories of gravity is \cite{Hayashi01091980, PhysRevD.80.104031}
\bqn
\lb{LA}
{\cal{L}}_{A}&=&-\frac{1}{4}\Big(c_1 F_{\mu\nu ij}e^{i \mu}e^{j \nu} + c_2 F_{\mu\nu ij}F^{\mu \sigma ik}e^{j \nu}e_{k \sigma}
+c_3 F_{\sigma \nu mj}F_{\mu \alpha in}e^{j \nu}e^{i \mu}e^{m \sigma}e^{n \alpha}\nb \\
&&~~~+c_4F_{\mu \nu ij}F^{\alpha \beta mn}e^{i \mu}e^j_{~\beta}e_{m \alpha}e_n^{~\nu}+c_5F_{\mu \nu ij}F^{\mu\nu ij}\Big),
\eqn
where
\bq
\lb{strength}
F_{\mu\nu ij}=\partial_{\nu}A_{ij\mu}-\partial_{\mu}A_{ij\nu}+A_{i~~\mu}^{~m}A_{mj\nu}-A_{i~~\nu}^{~m}A_{mj\mu}.
\eq
For the sake of generality this action is written for a general Riemann-Cartan geometry. Nevertheless, in this paper and the two preceding ones we solely work in a torsionless Riemannian space where one of the terms in the Lagrangian can be eliminated in terms of the rest. 

A well studied class of gauge theories of gravity is PGT. While the kinematics of these theories is well understood, their dynamics is subject to intense study even now. The first problem with understanding the dynamics of PGT is the existence of many invariant scalars made from curvature in comparison with a few of them in Yang-Mills theories. However, there is another more important difference between PGT and Yang-Mills theories. In the latter the dynamical variables are defined on the space-time, while in the former this is the space-time itself that is dynamical which is a direct consequence of the fact that in these theories, energy-momentum is one of the sources of gravity and this is the origin of problems like non-renormalizability. An alternative version of gauge theories of gravity is LGT in which metric is not dynamical and therefore energy-momentum is not a source of gravity.
The main difference between PGT and LGT is the variable with respect to which the variation of the Lagrangians is taken. To show the difference, the tetrad needs to be decomposed at each point of space-time
\bq
e_{i \mu}=\eta^{\bar{k} \bar{l}} e_{i \bar{k}}e_{\bar{l}\mu},
\eq
where bar refers to the frame being carried by the free falling observer at that point, and
\bqn
e_{\bar{l}\mu} &\equiv& \hat{e}_{\bar{l}}\cdot \hat{e}_{\mu}\nb\\
e_{i\bar{k}} &\equiv& \hat{e}_i \cdot \hat{e}_{\bar{k}},
\eqn
with $\hat{e}_i$, $\hat{e}_{\bar{l}}$, and $\hat{e}_{\mu}$ refer to unit vectors of the Minkowskian tangent space defined at that point, unit vectors carried by the free falling observer at that point, and the unit vectors tangent to coordinates at that point respectively. Therefore, the tetrad is broken into two components, one defined in the configuration space and the other defined in the Minkowskian tangent spaces. 
Any of the two components of the tetrad can be dynamical
\bqn
\lb{variation}
\delta e_{i\mu}=
\begin{cases}
\eta^{\bar{k} \bar{l}}e_{i \bar{k}}\delta e_{\bar{l} \mu} &  \text{Case I},\\
\eta^{\bar{k} \bar{l}}\delta e_{i \bar{k}}e_{\bar{l} \mu} &  \text{Case II}.
\end{cases}
\eqn
The first one holds in PGT while in LGT the second one is true. 
It is crucial to note that both of the theories are invariant under arbitrary change of coordinates in the configuration space and local Lorentz transformations in the tangent spaces and the latter equations only determine the variables with respect to which variations of Lagrangians are taken. Since every Lagrangian in a Riemannian space is made of the tetrad, metric, Christoffel symbols, and spin connections, it is important to see how each of these vary under the variations in equation \eqref{variation}. A variation in the metric comes from that in the tetrad section
\bqn
\delta g_{\mu\nu}=\eta^{ij}\left(e_{i\mu}\delta e_{j\nu}+\delta e_{i\mu} e_{j\nu}\right).
\eqn
Depending on which one of the equations in \eqref{variation} is substituted in this equation, the results will be different. If the first case is chosen, $\delta g_{\mu\nu}\neq 0$.  
On the other hand, if the second case is substituted, we end up with $\delta g_{\mu\nu}=0$ which can be easily seen if the followings are used
\bqn
\eta_{ij}&=&\eta^{\bar{i}\bar{j}}e_{i\bar{i}}e_{j\bar{j}},\nb\\
\delta \eta_{ij}&=&0,\nb\\
e_j^{~\bar{i}}\delta e_{i\bar{i}}&=&-e_i^{~\bar{i}}\delta e_{j\bar{i}}.
\eqn
An easier way to see why $\delta g_{\mu\nu}=0$ under the second case is to note that the metric is independent of how the Lorentz frames are set up in the Minkowskian tangent spaces at each point of space-time, i.e. the metric is independent of $e_{i\bar{i}}$ and can be written entirely in terms of $e_{\bar{i}\mu}$
\bqn
g_{\mu\nu}=\eta^{\bar{i}\bar{j}}e_{\bar{i}\mu}e_{\bar{j}\nu}.
\eqn 
Therefore, under the second case in equation \eqref{variation}, we can conclude that a variation in the metric, the metric compatible Christoffel symbols, the determinant of the tetrad, and the spin connections are
\bqn
\lb{deltag}
&&\delta g_{\mu \nu}=0,\nb\\
&&\delta \Gamma^{\alpha}_{\mu \nu}=0,\nb\\
&&\delta e=\delta \sqrt{-g}=0,\nb\\
&&\delta A_{ij\mu}=D_{\mu}\big(e_j^{~\nu}\delta e_{i \nu}\big),
\eqn
respectively and are all different from their corresponding ones in torsion-less PGT, i.e. if the first case in equation \eqref{variation} was chosen.
The two middle equations are easily justified knowing that a variation in the metric compatible Christoffel symbols, and also in the determinant of the metric can be entirely written in terms of $\delta g_{\mu \nu}$ which is itself zero. The very last equation is derived by taking a variation of the tetrad postulate, $D_{\mu}e_{i \nu}=0$,
\bqn
&&\partial_{\mu}\delta e_{i \nu}-\delta\Gamma^{\alpha}_{\mu \nu}e_{i \alpha}-\Gamma^{\alpha}_{\mu \nu}\delta e_{i \alpha}-\delta A_{ij \mu}e^j_{~\nu}-A_{ij \mu}\delta e^j_{~\nu}=\nb\\
&&D_{\mu}\delta e_{i \nu}-\delta\Gamma^{\alpha}_{\mu \nu}e_{i \alpha}-\delta A_{ij \mu}e^j_{~\nu}=0,
\eqn
and substituting $\delta \Gamma^{\alpha}_{\mu \nu}=0$ and the second case in equation \eqref{variation} in that
\bqn
\delta A_{ij\mu}=e_j^{~\nu}D_{\mu}\big(\eta^{\bar{k} \bar{l}}e_{\bar{l} \nu}\delta e_{i \bar{k}}\big).
\eqn

Before defining LGT through a Lagrangian it should be noted that since $\delta \Gamma^{\alpha}_{\mu \nu}$ and $\delta g_{\mu \nu}$ are zero in LGT, $\delta {\cal{L}}$ can be written in terms of $\delta A_{ij\mu}$ and $\delta e_{i\mu}$ that are dependent through the tetrad postulate. One way to impose this dependence when taking the variations is to use the last one of equation \eqref{deltag} and express $\delta A_{ij\mu}$ in terms of $\delta e_{i\mu}$. This ends up in a non-propagating constraint equation whose source term is zero. Another way of dealing with the tetrad postulate is to insert it as a constraint Lagrangian and treat $ A_{ij\mu}$ and $e_{i\mu}$ independently afterward.
Therefore, LGT is defined with
\bqn
\lb{action}
S&=&\int e d^4x\Big[{\cal{L}}_{A}+{\cal{L}}_{M}+{\cal{L}}_{C}\Big],
\eqn
with
\bqn
{\cal{L}}_{C}=S^{\mu \nu i}D_{\mu}e_{i \nu}, 
\eqn
where $S^{\mu \nu i}$ is a Lagrange multiplier, ${\cal{L}}_{M}$ is the Lagrangian of the standard model, and ${\cal{L}}_{A}$ is in general given by equation \eqref{LA}. However,
in order to remove some interactions that are not familiar from the standard model of particle physics, and are very likely to be non-renormalizable, we abandon four of the terms and restrict ourselves to the following
\bqn
\lb{NLA}
{\cal{L}}_{A}&=&-\frac{1}{4}c_5F_{\mu \nu ij}F^{\mu\nu ij}.
\eqn
Variation with respect to $A_{ij\mu}$ leads to a dynamical equation while that with respect to $e_{i\mu}$ gives a constraint equation. After solving the latter and substituting the solution in the former, the field equations for classical problems read
\bqn
\lb{classicalFieldEq}
D_{\nu}F^{\mu\nu ij}&=&4\pi G \tilde{J}^{\mu ij},
\eqn
where
\bqn
\lb{generalsource}
\tilde{J}_{\mu ij}&=&D_j T_{\mu i}-D_i T_{\mu j},\nb\\
T_{i \mu}&=&\frac{\delta{\cal{L}}_M}{\delta e_{i \mu}}.
\eqn
In the end it should be noted that only fermionic matters contribute in the field equations. This can be understood by noting that the source $J$ is determined through a variation in the matter Lagrangian which can be generally written as
\bqn
\lb{generalVariation}
\delta{\cal{L}}=\frac{\partial \cal{L}}{\partial g_{\mu\nu}}\delta g_{\mu\nu}+
\frac{\partial \cal{L}}{\partial \Gamma^{\alpha}_{\mu\nu}}\delta \Gamma^{\alpha}_{\mu\nu}+
\frac{\partial \cal{L}}{\partial e_{i\mu}}\delta e_{i\mu}+
\frac{\partial \cal{L}}{\partial A_{ij\mu}}\delta A_{ij\mu},
\eqn
where all of the $\delta$ terms can be expressed in terms of $\delta e_{i\bar{j}}$. A bosonic Lagrangian can be entirely written in terms of the metric and its determinant and the Christoffel symbols. Therefore, $\frac{\partial \cal{L}}{\partial e_{i\mu}}$ and $\frac{\partial \cal{L}}{\partial A_{ij\mu}}$ are both zero.  On the other hand the metric and the Christoffel symbols remain unchanged when $e_{i\bar{j}}$, the component of the tetrad defined entirely in the Minkowskian tangent spaces, is varied. Hence, $\delta g_{\mu\nu}$, and $\delta \Gamma^{\alpha}_{\mu\nu}$ are also zero. Inserting all the pieces into equation \eqref{generalVariation}, the variation in the bosonic Lagrangian is  $\delta{\cal{L}}_{\text{Boson}}=0$. Since this variation is the source term of the field equations, we can conclude that bosonic fields do not contribute to the field equations. 
We would like to emphasize on a subtle point. One may wish to write the Christoffel symbols in the bosonic Lagrangians in term of the spin connections and the tetrad. This however would not change the results. It can be easily seen by writing the Christoffel symbols in term of the spin connections and the tetrad and taking a direct variation
\bqn
\Gamma^{\alpha}_{\mu\nu}&=&e^{i\alpha}\left( \partial_{\mu}e_{i\nu}-A_{ij\mu}e^j_{~\nu}\right),\nb\\
\delta\Gamma^{\alpha}_{\mu\nu}&=&\delta e^{i\alpha}\left( \partial_{\mu}e_{i\nu}-A_{ij\mu}e^j_{~\nu}\right)+e^{i\alpha}\left( \partial_{\mu}\delta e_{i\nu}-A_{ij\mu}\delta e^j_{~\nu}
-\delta A_{ij\mu} e^j_{~\nu}\right). 
\eqn
Now if we rewrite the first term and insert $\delta A_{ij\mu}=D_{\mu}\big(e_j^{~\nu}\delta e_{i \nu}\big)$ in the second term and expand the covariant derivative, the result will be
\bqn
\delta\Gamma^{\alpha}_{\mu\nu}&=&\delta e^{i\alpha}e_{i\beta}\Gamma^{\beta}_{\mu\nu}
+e^{i\alpha}\left( \partial_{\mu}\delta e_{i\nu}-A_{ij\mu}\delta e^j_{~\nu}
-\partial_{\mu}\delta e_{i\nu}+A_{ij\mu}\delta e^j_{~\nu}+\Gamma^{\beta}_{\mu\nu}\delta e_{i\beta}\right)\nb\\
&=&\Gamma^{\beta}_{\mu\nu} \delta \delta^{\alpha}_{\beta}\nb\\
&=&0.
\eqn
This means even if the Christoffel symbols are written in term of the spin connections and the tetrad, the variation of the whole combination is still zero.
On the other hand, fermionic Lagrangians can not be rearranged and written purely in terms of the metric and the Christoffel symbols because even if the latter two are entirely known, the tetrads that directly appear in such Lagrangians can only be determined with ambiguity. This is in fact a result of the freedom in changing the Lorentz frames in the tangent spaces, see \cite{WeinbergBookonGR} for a thorough description. 
For the very same reason, Lagrangian \eqref{NLA} can not in general be written in terms of the metric and its derivatives alone even if the torsion is assumed zero right from the beginning.
In theories like GR where coordinate objects like the metric represent gravity, the tetrad and the spin connections can be eliminated in terms of the metric and its derivatives by fixing the Lorentz frames in the tangent spaces without affecting the final results. Fixing the Lorentz frames will however bias the results in theories like LGT where gravity is represented by objects, the spin connections in our case, that belong both to the coordinate space and the Lorentz space.

\section{Homogeneous and isotropic universe: a zeroth order solution to the universe}

A homogeneous and isotropic space-time is represented by the following tetrad
\bqn
e_{i \mu}=
\begin{pmatrix}
1&~&~&~\\
~&\frac{a(t)}{\sqrt{1-kr^2}}&~&~\\
~&~&ra(t)&~\\
~&~&~&r\sin(\theta)a(t)
\end{pmatrix}.
\eqn

In \cite{LGTRepo} a Mathematica code is provided that computes all of the objects presented below. 
The Christoffel symbols, $\Gamma^{\lambda}_{\mu\nu}$, can be easily calculated
\begin{align}
\Gamma^0_{11}&=\frac{a\dot{a}}{1-kr^2} , & \Gamma^0_{22}&=r^2a\dot{a},& \Gamma^0_{33}&=r^2sin^2(\theta)a\dot{a}, \nb\\
\Gamma^1_{11}&=\frac{kr}{1-kr^2} , & \Gamma^1_{10}&=\frac{\dot{a}}{a},& \Gamma^1_{22}&=-r(1-kr^2), \nb\\
\Gamma^1_{33}&=-r(1-kr^2)sin^2(\theta), & \Gamma^2_{12}&=\frac{1}{r},& \Gamma^2_{20}&=\frac{\dot{a}}{a}, \nb\\
\Gamma^2_{33}&=-cos(\theta)\sin(\theta), & \Gamma^3_{13}&=\frac{1}{r},& \Gamma^3_{23}&=\frac{cos(\theta)}{\sin(\theta)}, \nb\\
\Gamma^3_{30}&=\frac{\dot{a}}{a},  \nb\\
\end{align}
where dot indicates derivative with respect to time. The spin connections $A_{ij\mu}$ are
\begin{align}
A_{122}&=\sqrt{1-kr^2},&  A_{133}&=\sqrt{1-kr^2}\sin(\theta), &  A_{101}&=\frac{\dot{a}}{\sqrt{1-kr^2}},\nb\\
A_{233}&=cos(\theta),& A_{202}&=r\dot{a},&  A_{303}&=r\sin(\theta)\dot{a},\nb\\
\end{align}
and the strength tensor $F_{\mu\nu ij}$ reads
\begin{align}
F_{1212}&=\frac{r(k+\dot{a}^2)}{\sqrt{1-kr^2}},&F_{1313}&=\frac{r\sin(\theta)(k+\dot{a}^2)}{\sqrt{1-kr^2}}, & F_{1010}&=\frac{\ddot{a}}{\sqrt{1-kr^2}},\nb\\
F_{2323}&=r^2\sin(\theta)(k+\dot{a}^2), & F_{2020}&=r\ddot{a}, & F_{3030}&=r\sin(\theta)\ddot{a}.
\end{align} 
The only independent non-zero component of the source given in equation \eqref{generalsource} is
\bqn
\lb{SourceGeneral}
\tilde{J}^{rrt}&=&\sqrt{1-kr^2}\Big(\frac{(\rho_{_f}+p_{_f})\dot{a}+a\dot{p}_{_f}}{a^2}\Big),
\eqn
where the subscript $f$ refers to fermions because bosonic fields do not contribute in the source, as was explained in section \ref{LGTsec}. Also, we have used the following energy-momentum tensor for fermions
\bqn
\lb{energymomentum}
T_{_f\mu}^{~~\nu}=
\begin{pmatrix}
-\rho_{_f}&~&~&~\\
~&p_{_f}&~&~\\
~&~&p_{_f}&~\\
~&~&~&p_{_f}
\end{pmatrix}.
\eqn
The source given here can be written in a more applicable way if we use the conservation of energy-momentum, $D(T_{_{\text{fermion}}}+T_{_{\text{boson}}})=0$, which implies that
\bqn
\lb{ConservationOfEnergy}
\dot{\rho}_{_f}=-\frac{3\dot{a}}{a}\big(p_{_f}+\rho_{_f}\big)+G^0,
\eqn
with
\bqn
G^{\nu}=D_{\mu}T^{\mu\nu}_{\text{fermion}}=-D_{\mu}T^{\mu\nu}_{\text{boson}}.
\eqn 
Substituting equation \eqref{ConservationOfEnergy} into equation \eqref{SourceGeneral} results in the desired form of the source field
\bqn
\lb{JrrtLast}
\tilde{J}^{rrt}=\frac{\sqrt{1-kr^2}}{a^2}\Big( (1-3w)(1+w)\rho_{_f}\dot{a}+awG^{0}+a\dot{w}\rho_{_f} \Big),
\eqn
where $w=\frac{p_f}{\rho_f}$.
Appearance of $G^0$ in this equation seems contradictory with our previous statement that bosonic fields do not enter the field equations. However, it should be noted that $G^{\nu}=-D_{\mu}T^{\mu\nu}_{\text{boson}}=D_{\mu}T^{\mu\nu}_{\text{fermion}}=0$ if matter does not transform from fermion to boson, or vice versa. Therefore, $G^0$ is indeed a change in the fermionic content of the system through creation or annihilation which is accompanied by annihilation or creation of bosons. 

In the end, a direct substitution shows that $D_{\mu}\tilde{J}^{\mu ij}=0$ is trivially satisfied with no further constraint. After substituting everything into field equation \eqref{classicalFieldEq}, it reads  
\bqn
\lb{cosmologyFieldEq}
\dddot{a}=2\frac{\dot{a}^3}{a^2}-\frac{\dot{a}\ddot{a}}{a}+2k\frac{\dot{a}}{a^2}+4\pi G a^2 \frac{\tilde{J}^{rrt}}{\sqrt{1-kr^2}}.
\eqn
Here, unlike in GR, sum of energy densities from fermions and bosons at present time is not limited to a given value because bosons do not contribute to the field equations. However, through equation \eqref{cosmologyFieldEq} we can still find a relation for the sum of the fermionic energy in the universe at present time. 
In GR the bound on energy content of the universe leads to a value for vacuum energy that contradicts the one given by field theory while in LGT, whatever the value of the vacuum energy is, there will be no contradiction since the vacuum energy does not enter equation \eqref{cosmologyFieldEq}. The latter statement holds if the vacuum energy is being described by ${\cal{L}}_{\text{vacuum}}=\sqrt{-g}\times\text{constant}$ as is given in \cite{RevModPhys.61.1}.
While the vacuum energy predicted by field theory is often described by latter Lagrangian, if a fermionic field with $w=-1$ like those suggested in \cite{Grams2014} exist, it can contribute in the second term of the source \eqref{JrrtLast} but not in the first and third terms due to the equation of state of the field. Therefore, even if such a field exist, the contribution is nonzero only when the field is created or annihilated, i.e. when $G^0$ is not zero.
 
A simple evaluation of the differential equation above shows that in an expanding flat universe the time derivative of acceleration is positive for a negligible matter and a negative acceleration. Therefore, some time after the matter content of the universe is diluted enough, any decelerating expansion will spontaneously transit to an accelerating one. 

Equation \eqref{cosmologyFieldEq} is not convenient for exploring early times in the universe. It is instead easier if physical quantities are expressed in terms of the scale factor. 
Therefore, a simple change of variable will be very helpful. If we show $\frac{d}{da}$ with a prime and rename $\dot{a}$ to $y$, equation \eqref{cosmologyFieldEq} can be rewritten as a second order differential equation
\bqn
\lb{cosmologyFieldEq1}
&&(ayy')'-2\frac{y^2}{a}=4\pi G a^3 \frac{\tilde{J}^{rrt}}{y},\nb\\
&&\tilde{J}^{rrt}=\frac{1}{a^2} \Big( (1-3w)(1+w)\rho_{_f} y+awG^{0}+a\rho_{_f} yw'  \Big),
\eqn
where $k=0$ is now applied. At this point we would like to discuss the three terms in the source $\tilde{J}^{rrt}$. We first note that if $G^{0}=0$, equation \eqref{ConservationOfEnergy} can be solved to find the density in terms of the scale factor
\bqn
\rho_{_f} =  \frac{\rho_{{_f}0}}{a^{3(1+w)}},
\eqn
where the naught refers to present time and also $a_0=1$ is assumed. The first term in $\tilde{J}^{rrt}$ is zero for relativistic matters with $w=\frac{1}{3}$ while for non-relativistic matters, it takes the following form
\bqn
\lb{nonrelSource}
\tilde{J}_{\text{first}}^{rrt}=\frac{\rho_{M_0}}{a^5} y.
\eqn 
The second term is non-zero when a fermionic matter is created or annihilated in the universe. Since we are concerned with timescales much longer than that of pair production, we choose to model the process with a step function
\bqn
\rho_{_f} = \rho_{_f\text{initial}} + \Delta \rho_{_f} \Theta(a-\tilde{a}),
\eqn
where $\Delta \rho_{_f}$ is assumed to be constant and $\tilde{a}$ is the moment at which such a transition happens. 
$G^0$ can be determined if a time derivative of this equation is taken and then compared with conservation law \eqref{ConservationOfEnergy},
\bqn
\rho_{_f}'&=&\rho_{_f\text{initial}}'+\Delta \rho_{_f} \delta(a-\tilde{a})\nb\\
&=&-\frac{3}{a}\big(p_{_f}+\rho_{_f}\big)+\frac{G^0}{y}.
\eqn
If no matter was created or annihilated, the conservation law would be $\rho_{\text{initial}}'=-\frac{3}{a}\big(p+\rho\big)$. Therefore
\bqn
G^0=y\Delta \rho_{_f} \delta(a-\tilde{a}),
\eqn
and the source term gets the following form
\bqn
\lb{secondsource}
\tilde{J}_{\text{second}}^{rrt}&=&\frac{w y}{a} \Delta \rho_{_f} \delta(a-\tilde{a}).
\eqn
To get a sense of this source, one can imagine a plasma of relativistic electrons, positrons, and photons. Since the matter is relativistic, its creation and annihilation has the same rate, $\sum \Delta \rho_{_f} \sim 0$. When the universe expands enough, particles loose their energy and become non-relativistic. At this time the whole fermionic matter annihilates. But, now $w \sim 0$ and again the source is negligible. We may conclude that the term is only important at the beginning of the universe when matter was created. But, as the numerical study below shows, at the beginning of time and assuming only the known matter, the magnitude of this source is negligible comparing with the geometrical terms. Therefore, unless an unknown fermionic matter is created or annihilated, probably with $w=-1$ or $w=\frac{1}{3}$ such that it does not contribute to the other source terms, this source can be totally neglected. 

\begin{figure}[tbp]
\centering
\includegraphics[width=14cm]{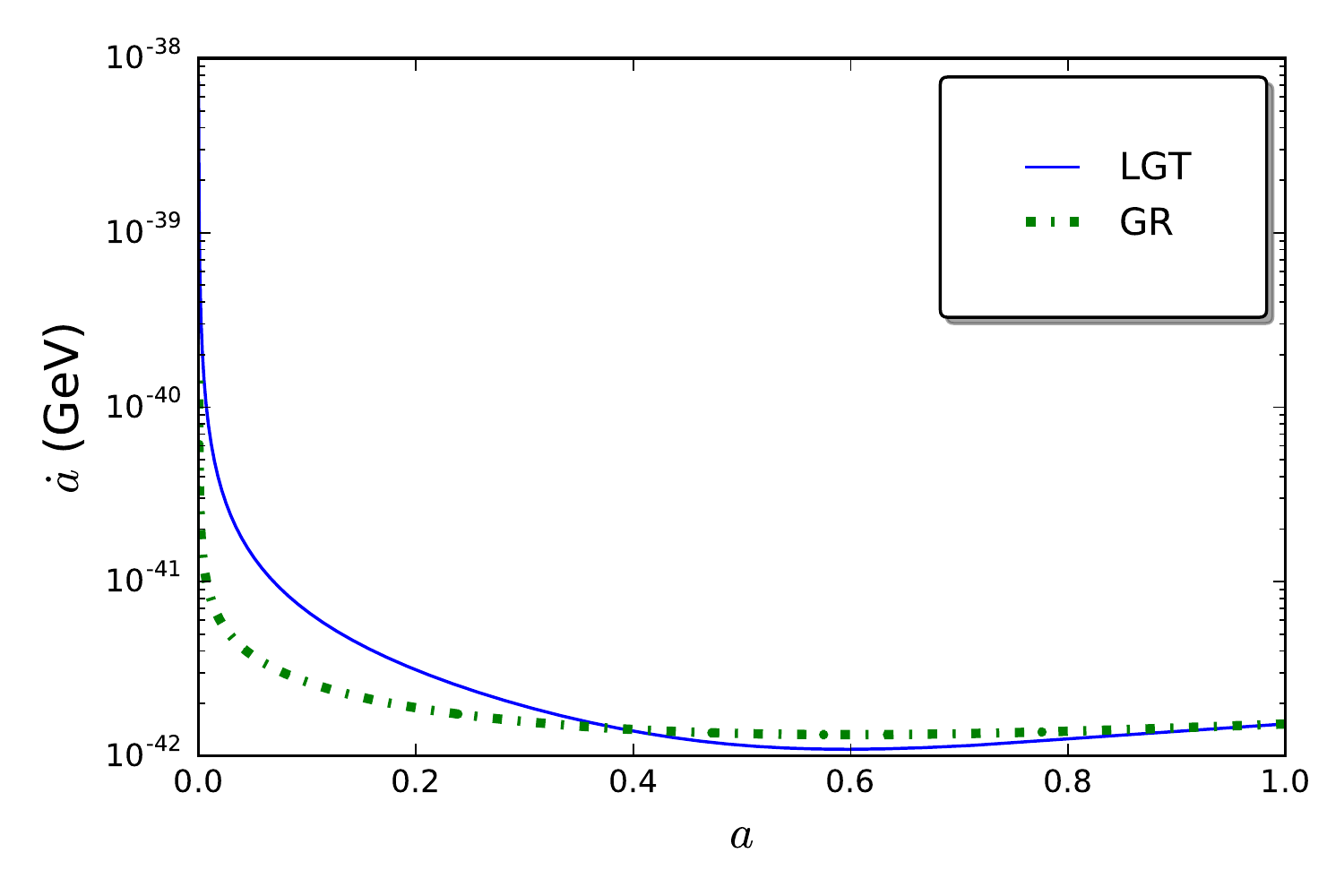}
\caption{The time derivative of the scale factor $\frac{da}{dt}$ versus the scale factor $a$. The acceleration $\ddot{a}$ of the universe change sign at around $a \sim 0.6$.}
\lb{timeDer_scale}
\vspace{0.75cm}
\end{figure}

The third term in $\tilde{J}^{rrt}$ is non-zero when a relativistic matter looses its energy and becomes non-relativistic while still in thermal equilibrium with the rest of matter. The reason for the latter condition is because otherwise matter's distribution function, and therefore its equation of state, remains the same at later times even when it is non-relativistic. Consequently, the term can be non-zero in early times only. The transition can be crudely modeled by $w=\frac{1}{3}\Theta(\tilde{a}-a)$
and consequently
\bqn
w'=-\frac{1}{3}\delta(\tilde{a}-a). 
\eqn
The third source term therefore is 
\bqn
\lb{thirdsource}
\tilde{J}_{\text{third}}^{rrt}&=&\frac{-\rho_{_f} y}{3a}  \delta(a-\tilde{a}).
\eqn
Our numerical study shows that this term can be safely ignored even at early times due to its small magnitude.

In the following we will numerically solve the differential equation in order to get a description of the history of the universe, but before that we would like to find the exact solutions for when $\tilde{J}^{rrt}=0$. It should be noted that the latter condition can be satisfied in two ways. First, when there is no matter at all. Second, when a matter exist with an equation of state $w=\frac{1}{3}$ or $w=-1$ provided there is also no particle creation or annihilation. Substitution of $y=ca^n$ in the differential equation results in the following equation
\bqn
n^2-1=0.
\eqn
Consequently, the two exact solutions for when $\tilde{J}^{rrt}=0$ are $y=ca^{\pm 1}$ with any arbitrary $c$. The Hubble parameter for these two is 
\bqn
\lb{vacuumsolution}
H=
\begin{cases}
\frac{c}{a^2},&\\
c~. & \\
\end{cases}
\eqn 
The two solutions are already known. The latter is a de Sitter space which is also a solution of GR if a dark energy is assumed while the former represents a solution of GR when universe is filled with a relativistic matter. Later by a numerical analysis we show that in LGT the universe expands like the first exact solution, $H=\frac{c}{a^2}$, in the early times and like the other one, $H=c$, in the late times. This is expected since in both of the mentioned periods $\tilde{J}^{rrt}$ is negligible. 

In the end we would like to discuss how an inflationary era in the very beginning of the universe can be addressed in LGT. We first need to show that in the very beginning of the universe there has been an accelerating expansion. Next, we not only need to show that it can be stopped, but also show that the expansion of the universe after the inflation is consistent with that obtained from our numerical study below. 
Also, we expect that when the inflation is being stopped, current matter in the universe is created. To meet the mentioned expectations, we first need the Hubble parameter to be $H=c_1=\text{constant}$ in the inflationary era. On the other hand, the numerical study presented below indicates that at early times, but after inflation, the universe should expand as $H= c_2a^{-2}$ where $c_2$ is a constant. We note that both of the needed expansion modes are already solutions of LGT when $\tilde{J}^{rrt}=0$ and are given in equation \eqref{vacuumsolution}. It also should be emphasized that in LGT, unlike in GR where the potential of inflaton should satisfy some conditions \cite{Linde1983,Silverstein2004,Cervantes-Cota1995}, the inflationary solution holds as far as $\tilde{J}^{rrt}=0$ and is not driven by fields. At this point, all we need is to first see if the condition $\tilde{J}^{rrt}=0$ can be satisfied both during and after the inflationary era and second see if there is a way to connect the two solutions at the boundary. We first study the boundary conditions. According to equation \eqref{cosmologyFieldEq1} 
\bqn
ayy' |_{a_+} - ayy' |_{a_-}=\frac{4\pi G a^3}{y}\int_{a_-}^{a_+} \tilde{J}^{rrt} da,
\eqn
where plus and minus subscripts of $a$ refer to after and during inflation respectively. Moreover, both $a$ and $y$ are continuous. Therefore, at the boundary
\bqn
a_+&=&a_-,\nb\\
c_1&=&c_2 a_+^{-2},\nb\\
\int_{a_-}^{a_+} \tilde{J}^{rrt} da&=&-\frac{c_1^3}{2\pi G},
\eqn
where $c_2$ is fixed by the numerical study below. For the last equation to hold, $\tilde{J}^{rrt}$ needs to be of a delta function type. Also, the matter in the universe should be created at this time. The source associated with particle creation and annihilation which is also in the form of Dirac delta function is given in equation \eqref{secondsource}. After substituting that in the integral above
\bqn
w \Delta \rho_f = -\frac{c_1^2}{2\pi G}, 
\eqn
which is satisfied only if the left hand side is negative. We can imagine two scenarios that satisfy all the requirements mentioned above. Whether any of them can be supported by physical principles is left for future works. The first scenario is that during the inflationary era there is no matter at all, then suddenly a fermionic matter with $w=-1$ is created, i.e. $\Delta \rho_f > 0$. This scenario is less likely because a field is created out of nothing. Another scenario is that a relativistic fermionic matter existed during the inflationary era but suddenly annihilated, i.e. $\Delta \rho_f < 0$, to mostly bosons, which do not contribute to the field equations and therefore to the expansion of the universe, together with a bit of fermions as seeds for current matter in the universe. This interaction can be represented by ${\cal{L}}_{\text{int}}=\bar{\Psi}\Psi\Phi$. It should be noted that, form of this interaction is the same as the one in the standard model which couples the Higgs field with fermions. However, whether $\Phi$ in the interaction above can be the Higgs field is a question left to future work.

\subsection{Numerical Solution}
We can use numerical analysis to get a thorough picture of the evolution of the universe from the beginning to the end. This is done by setting the initial values at present time based on the current observations. Next, the differential equation in \eqref{cosmologyFieldEq1}, and only that, determines how the universe looked like in the past.  
The first term in $\tilde{J}^{rrt}$ is zero for relativistic matters with $w=\frac{1}{3}$ while for non-relativistic matters, it is given by equation \eqref{nonrelSource}. Therefore, the source is known if we determine the density of cold matter in the universe at present time. This is assumed to be $\rho_{M_0}=0.3 \rho_{\text{c}}$ where the critical energy density is defined by
\bqn
\rho_{\text{c}}=\frac{3H_0^2}{8\pi G}.
\eqn
As was discussed above, if we do not assume any unknown matter, the other two terms of $\tilde{J}^{rrt}$ in \eqref{cosmologyFieldEq1} are either exactly zero or negligible and therefore will be ignored in our numerical study. 
\begin{table}[th]
\caption{
Initial conditions used for numerically solving the field equation.
}
\begin{center}
\begin{tabular}{c|cccccc}
\hline
Quantity       & ~~$a_0$ & ~~$H_0$~~ & $\Omega_{M_0}$~ & $q_0$ ~ & $k$   \\
\hline
Value at $t=0$ & ~~1   & ~~71~~ & 0.3~ & -1 ~ & 0  \\
\hline
\end{tabular}
\end{center}
\label{tableinitialvalue}
\end{table}
For the Hubble parameter at present time, $71\times (2 \times 10^{-44}~\text{GeV})$ is assumed in accordance with observations. 
To set the initial values for $y_0$ and $y_0'$, the following two equations are used
\bqn
y_0&=&H_0,\nb\\
y_0'&=&-q_0\times H_0 ,
\eqn
where $q_0$ is the deceleration parameter at present time and is taken to be $-1$. Also, value of $y_0''$ is known through equation \eqref{cosmologyFieldEq1}. The assumed initial values are all gathered in table \ref{tableinitialvalue}. It should be emphasized that because of their equation of state, relativistic matters do not contribute to equation \eqref{cosmologyFieldEq1} at even early times and therefore their energy density at present time does not affect the expansion of the universe and for this reason number of relativistic particles are not limited in LGT. 
At this point the fourth order Runge-Kutta method can be used to numerically solve the differential equation. A python code that does the numerical evaluations and also makes the plots is provided in \cite{LGTRepo}.

\begin{figure}[tbp]
\centering
\includegraphics[width=14cm]{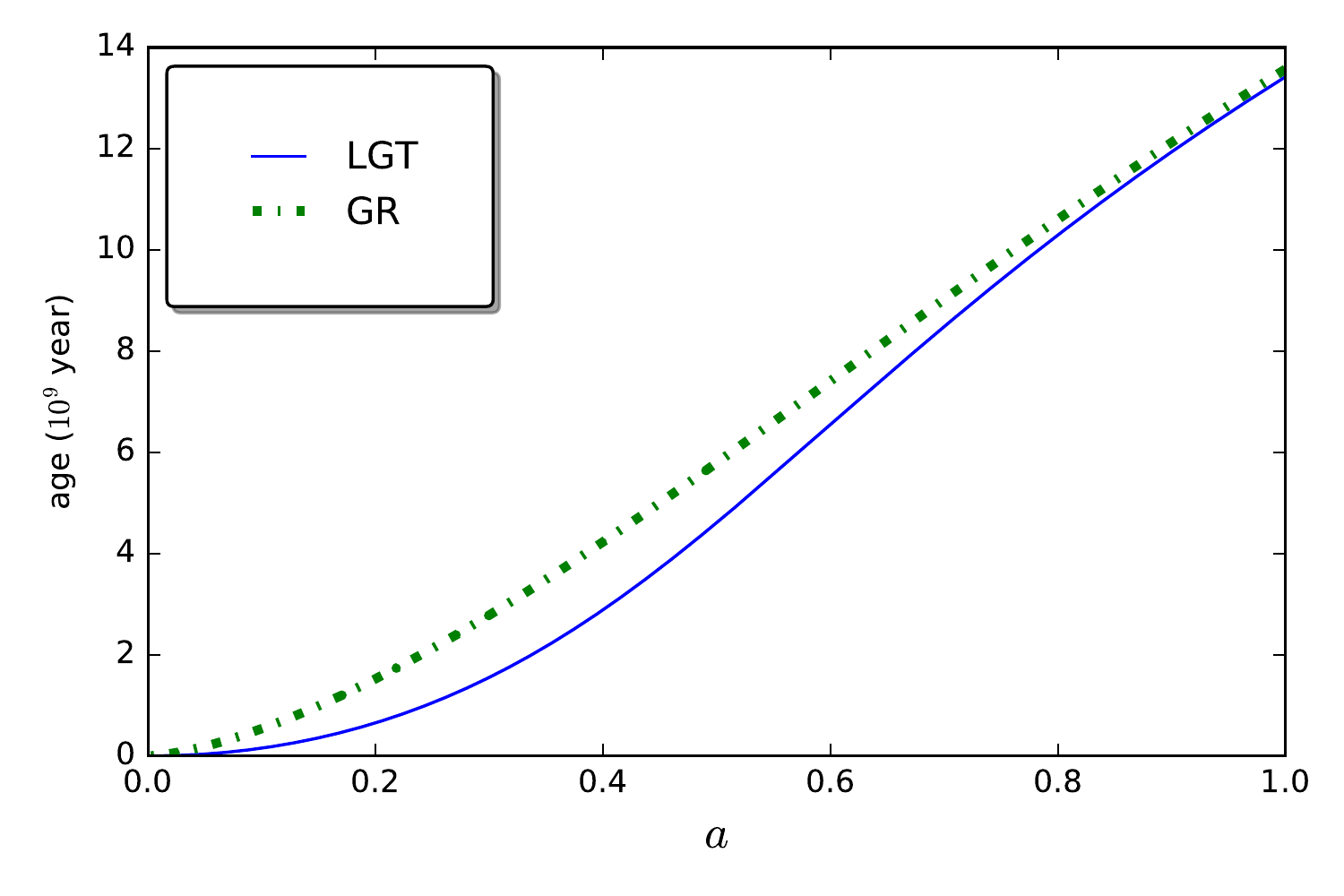}
\caption{Predictions of LGT and GR for the age of the universe versus the scale factor.}
\lb{age_scale}
\vspace{0.75cm}
\end{figure}

\begin{figure}[tbp]
\centering
\includegraphics[width=12cm]{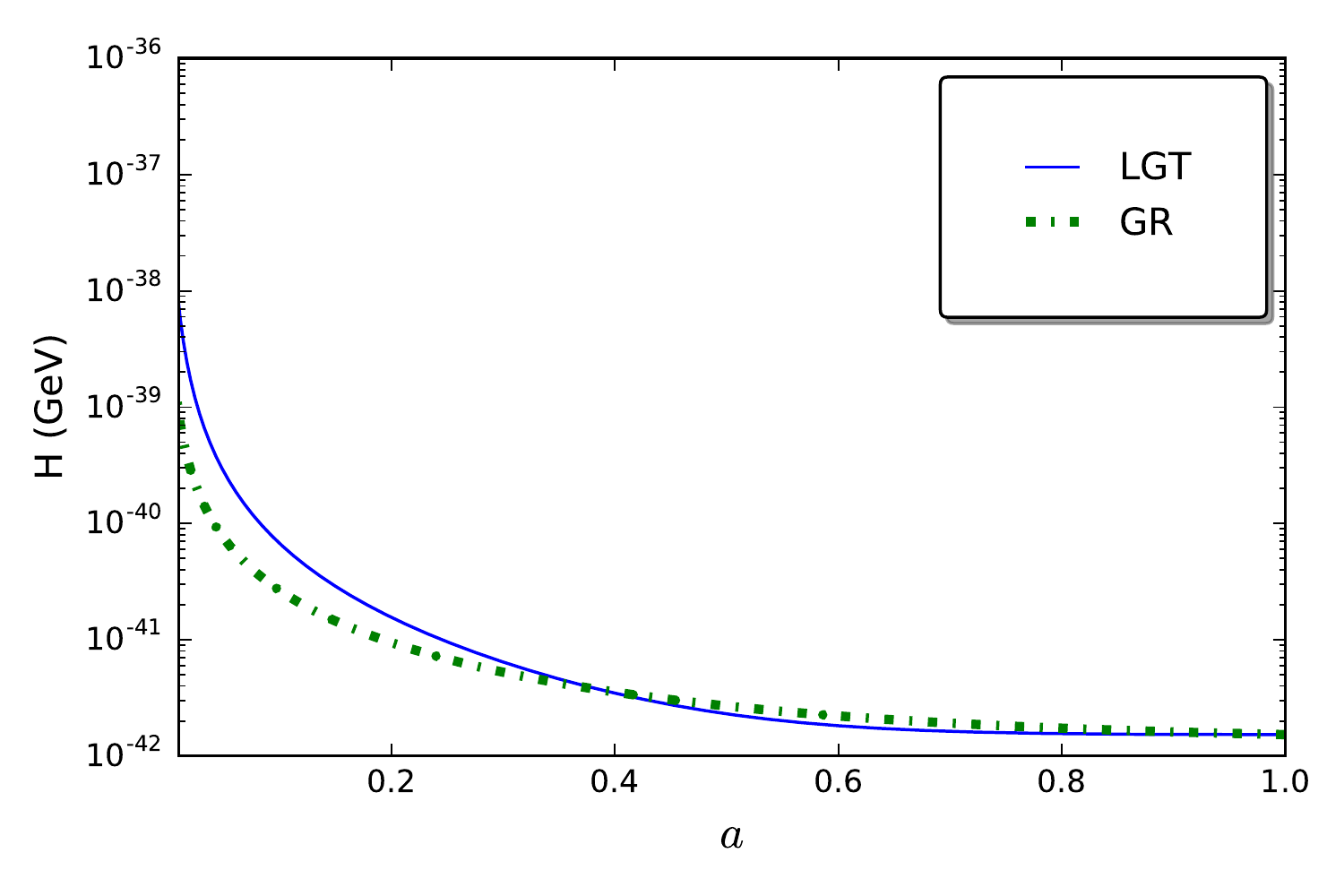}
\includegraphics[width=12cm]{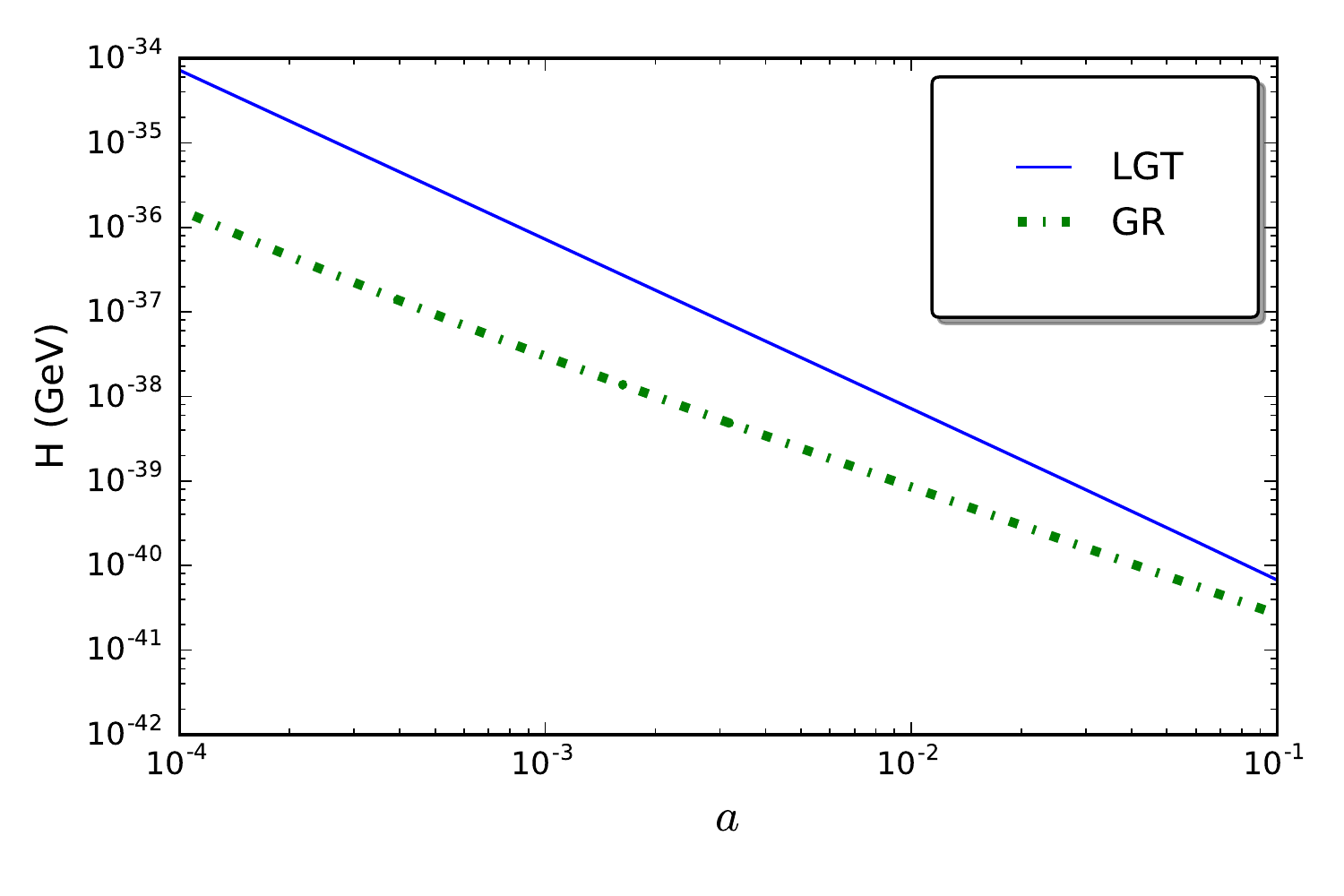}
\includegraphics[width=12cm]{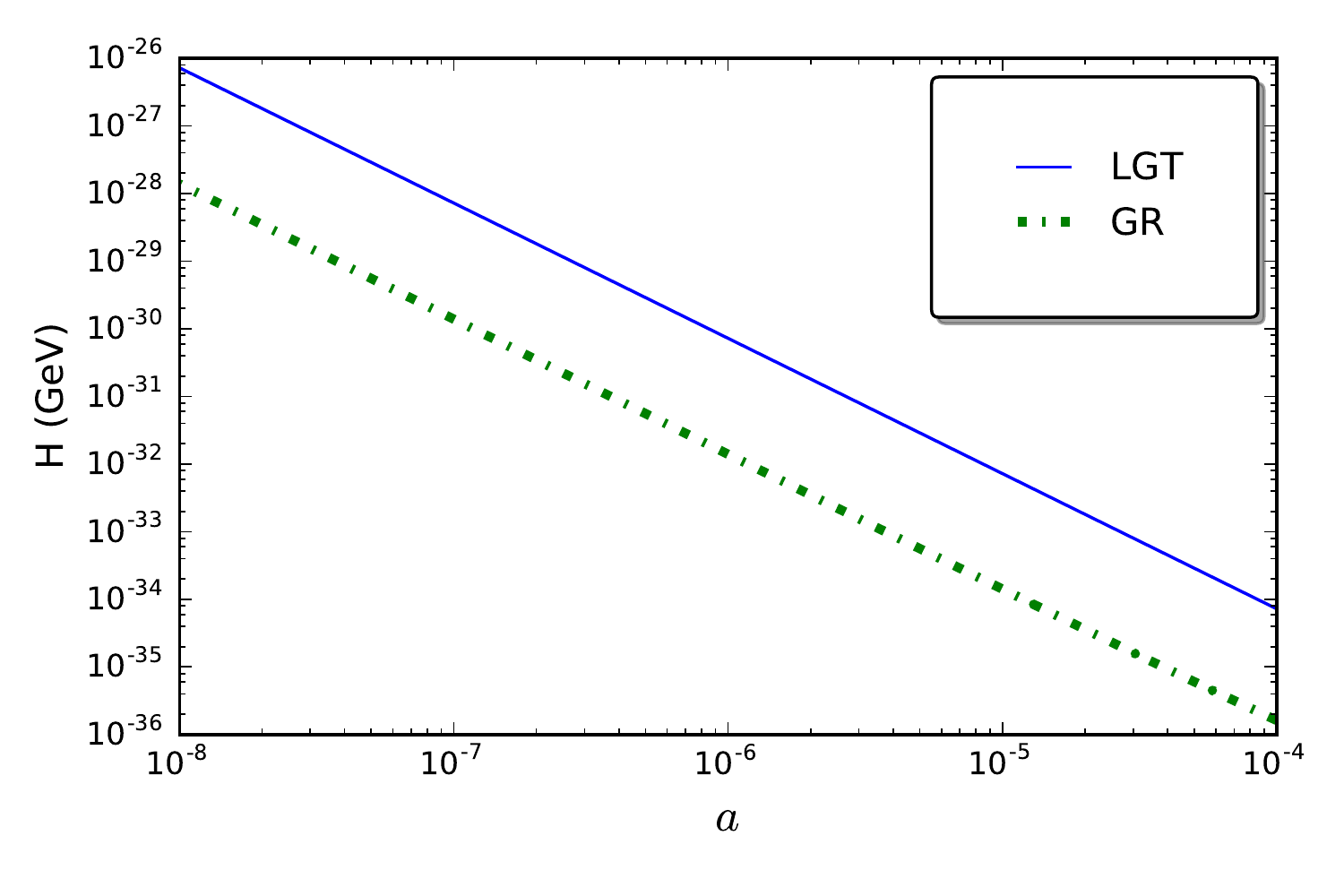}
\caption{Hubble parameter versus the scale factor in three different stages.}
\lb{hubble_scale1}
\vspace{0.75cm}
\end{figure}

\begin{figure}[tbp]
\centering
\includegraphics[width=14cm]{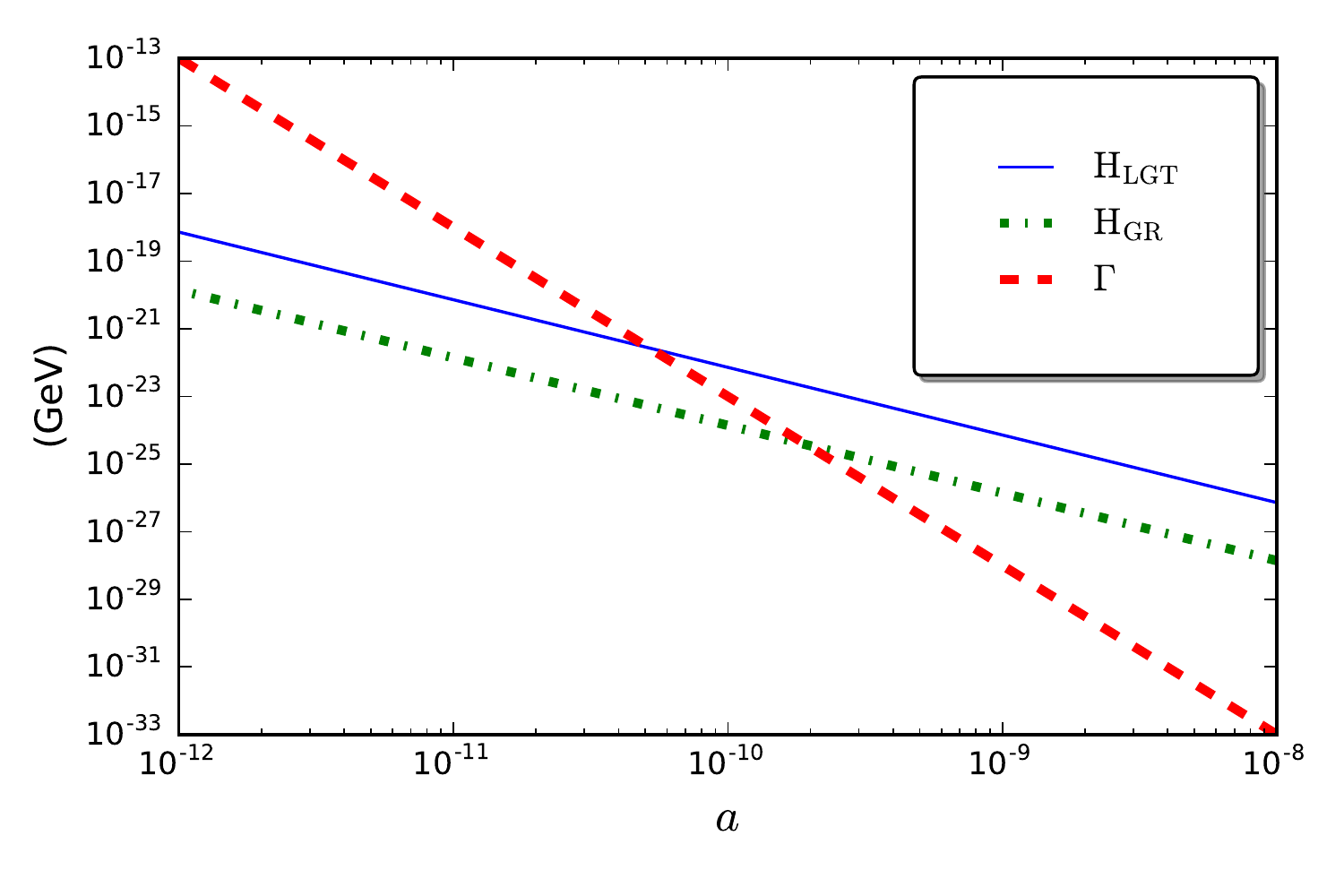}
\caption{Inverse of neutron free time $\Gamma$ and Hubble parameter from LGT and GR as a function of the scale factor. Neutrons freeze out when the Hubble parameter is roughly equal to the inverse of neutrons free time. This plot indicates that the freeze out temperature in LGT is higher than that in GR.}
\lb{hubble_scale2}
\vspace{0.75cm}
\end{figure} 

In figure \ref{timeDer_scale} the time derivative of the scale factor versus the scale factor, where the slope is $\frac{\ddot{a}}{\dot{a}}$, is plotted. Since the slope's denominator is always positive, the plot indicates that $\ddot{a}$ is 
negative at early times but becomes positive in late times. The transition from deceleration to acceleration happens spontaneously without assuming dark energy at $a \sim 0.6$ which is roughly at the same time as predicted by GR. 
The age of the universe as a function of the scale factor predicted by both LGT and GR are shown in figure \ref{age_scale}. Although in the matter dominated era the GR universe is older than the LGT one, at present time both of the theories predict almost the same ages for the universe. 
The Hubble parameter as a function of the scale factor is given in figures \ref{hubble_scale1} and \ref{hubble_scale2} in four different stages. In GR the universe is dominated by radiation when $a \lesssim 10^{-4}$, starting from the bottom plot in figure \ref{hubble_scale1}. On the other hand, as this plot as well as the one in figure \ref{hubble_scale2} show,
LGT predicts the same expansion $H \propto a^{-2}$ during this era, i.e. the curves predicted by both theories are parallel. Therefore, we can conclude that the transition from the matter to radiation domination is predicted by LGT to be at the same time as in GR. 
As was discussed above, $H \propto a^{-2}$ is LGT's exact solution when $\tilde{J}^{rrt}=0$ which is the case for relativistic matters. It is important to note that in GR this expansion is very sensitive to the magnitude of the relativistic matter at that time and this imposes stringent limits on the number of relativistic particles \cite{Steigman1977,Steigman2012,Abazajian2001}.
In LGT there is no such limit because $\tilde{J}^{rrt}=0$ for relativistic matters and the expansion remains the same no matter how many relativistic particles exist. 

Observation of light element abundances indicate that the baryonic matter in the universe is made of 75\% hydrogen, 25\% helium and a negligible amount of other light elements. The values depend very much on the expansion rate of the universe during the radiation dominated era as well as neutrino's chemical potential \cite{Dolgov2002,Hansen2002}. In the early universe at temperatures higher than a few MeV, protons and neutrons existed in nearly the same amount and interacted through the following weak interactions
\bqn
&&p+e^- \leftrightarrow n+\nu_e,\nb\\
&&n+e^+ \leftrightarrow p+\bar{\nu}_e,\nb\\
&&n \leftrightarrow p + e^- + \bar{\nu}_e.
\eqn
The rate of these interactions is crudely calculated to be \cite{Bernstein1989,Cyburt2016} 
\bqn
\lb{neutron_protonInt}
\Gamma = 1.2~G_F^2~T^5,
\eqn
where $G_{\text{F}}$ is the Fermi coupling constant. The interactions above stop when this rate is equal to the expansion rate of the universe
\bqn
H \sim \Gamma,
\eqn
when all the existing neutrons freeze out. To find out the temperature when this happens, we have plotted both $H$ and $\Gamma$ in figure \ref{hubble_scale2}. The plot indicates that the freeze-out temperature in GR is 1.2 MeV while in LGT it is 4.4 MeV. On the other hand, the ratio of neutron number density to that of protons is determined by the Boltzmann equation 
\bqn
\lb{neutron_proton_ratio}
\frac{n_n}{n_p}=\text{exp}\left(-\frac{m_n-m_p+\mu_{\nu_e}-\mu_{e}}{T}\right),
\eqn
where $m_n-m_p\sim 1.3$~MeV is the difference in neutron and proton mass, while $\mu_e$ and $\mu_{\nu_e}$ are the chemical potentials of electron and its neutrino respectively. 
In GR the ratio is consistent with the observations if the chemical potentials, i.e. lepton asymmetries, are negligible \cite{Kang1992,Hamann2008}.
The higher neutron freeze out temperature in LGT can be crudely interpreted as non-negligible lepton asymmetry. However, we postpone a definitive conclusion until after a derivation of light element abundances which is left for future work.
Since the universe is electrically neutral, the lepton asymmetry can only lie in the neutrino section \cite{Mukhanov_}. In the standard model, neutrinos are assumed massless. However, recent observations indicate that they are oscillating between their different modes. This itself is an indication that neutrinos are massive and therefore necessitates exploration of physics beyond the standard model \cite{Super-Kamiokande}. Although not experimentally verified, it is widely argued that neutrinos are Majorana fermions in which case we should observe a larger lepton asymmetry than predicted by the standard model. For a review of the subject see \cite{Davidson_}. 
It should be also mentioned that current observational constraints on lepton asymmetry is very weak and direct measurement of lepton asymmetry is not possible with current detectors because of non-relativistic nature of primordial neutrinos at present time and their extremely weak interactions.

\section{Conclusions}

A homogeneous and isotropic universe has been studied within Lorentz gauge theory of gravity. Two exact solutions are derived for when the source of the theory is zero. One of the solutions is the de Sitter space-time which is also GR solution if vacuum energy is assumed while the other is the same as the GR solution for a radiation dominated universe. 
It has been shown that, unlike in GR, there is no constraint on the sum of the total, fermionic plus bosonic, energy content of the universe. Therefore, the extremely huge value of the vacuum energy predicted by field theory does not contradict LGT's predictions. This is because, as is discussed in \cite{Borzou2016_2}, vacuum energy does not enter LGT's field equations and consequently does not contribute to the expansion of the universe. 

Through a numerical study the expansion of the universe is studied. The initial values have been chosen based on the current observations. Age of the universe has been found to be $\sim 13.4$ billion years which is very close to the currently accepted value. It also has been shown that the universe spontaneously transits from a decelerating to an accelerating expansion at redshift $\sim 0.6$ comparable to that in GR with the difference that in LGT the transition is driven by geometrical terms. In early times when $ a < 10^{-4}$, LGT's expansion rate has the same form as that in GR, i.e. $H\propto \frac{1}{a^2}$. This means that in LGT the transition to the radiation dominated universe takes place at the same time as predicted by GR. It is also shown that neutrons freeze out at higher temperature in LGT than the temperature derived in GR. A large lepton asymmetry in LGT would be the preliminary explanation. However, a definitive conclusion is postponed until after a detailed study of light element abundances is carried out. 

It is demonstrated that an inflationary period in the beginning of time can be addressed in LGT. The inflation is driven by geometrical terms and not by a field and therefore there is no need for an inflaton with slow-roll or similar conditions.

~\\{\bf Acknowledgements:}
We are pleased to thank the referees for their useful comments and suggestions.

\end{document}